\documentclass[twocolumn,aps,showpacs,superscriptaddress]{revtex4}
\usepackage[latin9]{inputenc}
\usepackage{amsmath}
\usepackage{amssymb}
\usepackage{graphicx}
\usepackage{esint}

\makeatletter
\@ifundefined{textcolor}{}
{%
 \definecolor{BLACK}{gray}{0}
 \definecolor{WHITE}{gray}{1}
 \definecolor{RED}{rgb}{1,0,0}
 \definecolor{GREEN}{rgb}{0,1,0}
 \definecolor{BLUE}{rgb}{0,0,1}
 \definecolor{CYAN}{cmyk}{1,0,0,0}
 \definecolor{MAGENTA}{cmyk}{0,1,0,0}
 \definecolor{YELLOW}{cmyk}{0,0,1,0}
}

\usepackage{color}

\newcommand{\beq}{\begin{equation}}
\newcommand{\eeq}{\end{equation}}
\newcommand{\bea}{\begin{eqnarray}}
\newcommand{\eea}{\end{eqnarray}}

\makeatother

\begin{document}

\title{Displacement and annihilation of Dirac gap-nodes in $d$-wave iron-based
superconductors}

\author{Andrey V.~Chubukov}

\affiliation{School of Physics and Astronomy, University of Minnesota, Minneapolis,
MN 55455, USA}

\author{Oskar Vafek}

\affiliation{Department of Physics and National High Magnetic Field Laboratory,
Florida State University, Tallahassee, Florida 32306 USA}

\author{Rafael M. Fernandes}

\affiliation{School of Physics and Astronomy, University of Minnesota, Minneapolis,
MN 55455, USA}

\date{\today}

\begin{abstract}
Several experimental and theoretical arguments have been made in favor
of a $d-$wave symmetry for the superconducting state in some Fe-based
materials. It is a common belief that a $d-$wave gap in the Fe-based
superconductors must have nodes on the Fermi surfaces centered at
the $\Gamma$ point of the Brillouin zone. Here we show that, while
this is the case for a single Fermi surface made out of a single orbital,
the situation is more complex if there is an even number of Fermi
surfaces made out of different orbitals. In particular, we show that
for the two $\Gamma$-centered hole Fermi surfaces made out of $d_{xz}$
and $d_{yz}$ orbitals, the nodal points still exist near $T_{c}$ along
the symmetry-imposed directions, but are are displaced to momenta between the two Fermi surfaces. 
 If the two hole pockets are close enough, pairs of nodal points can merge
and annihilate at some $T<T_{c}$, making the $d-$wave state completely
nodeless. These results imply that photoemission evidence for a nodeless
gap on the $d_{xz}/d_{yz}$ Fermi surfaces of KFe$_{2}$As$_{2}$
does not rule out $d-$wave gap symmetry in this material, while a
nodeless gap observed on the $d_{xy}$ pocket in K$_{x}$Fe$_{2-y}$Se$_{2}$
is truly inconsistent with the $d-$wave gap symmetry.
\end{abstract}
\maketitle

\section{Introduction}

One of the most interesting features of Fe-based superconductors (FeSC)
is the observation of different structures of the superconducting
(SC) gap in different materials, which may indicate that the gap symmetry
in FeSC is material dependent.~\cite{review} Weakly or moderately
doped FeSC have both hole and electron pockets, and the gap symmetry
there is very likely $s-$wave, with a $\pi$ phase shift between
hole pockets and electron pockets -- the so-called $s^{+-}$-wave
state~\cite{mazin}. The situation is less clear in materials with
only one type of Fermi pocket, such as strongly hole doped KFe$_{2}$As$_{2}$,
which contain only hole pockets~\cite{KFeAs}, and K$_{x}$Fe$_{2-y}$Se$_{2}$
or monolayer FeSe, which have only electron pockets~\cite{KFeSe}.
Thermal conductivity and Raman scattering measurements in KFe$_{2}$As$_{2}$~\cite{louis,louis_2,raman},
as well as the observation of a neutron resonance peak in the superconducting
state of K$_{x}$Fe$_{2-y}$Se$_{2}$ \cite{neutr_kfese}, were interpreted
as evidence for a $d-$wave gap symmetry in these materials. Theoretical
studies also found a strong enhancement of the $d-$wave superconducting
susceptibility~\cite{theory_kfe,laha}, and at least one study of
KFe$_{2}$As$_{2}$ have found~\cite{frg_thomale} a much stronger
attraction in the $d-$wave channel than in the $s^{+-}$ channel.

\begin{figure}[h]
\centering{}\includegraphics[width=0.9\columnwidth]{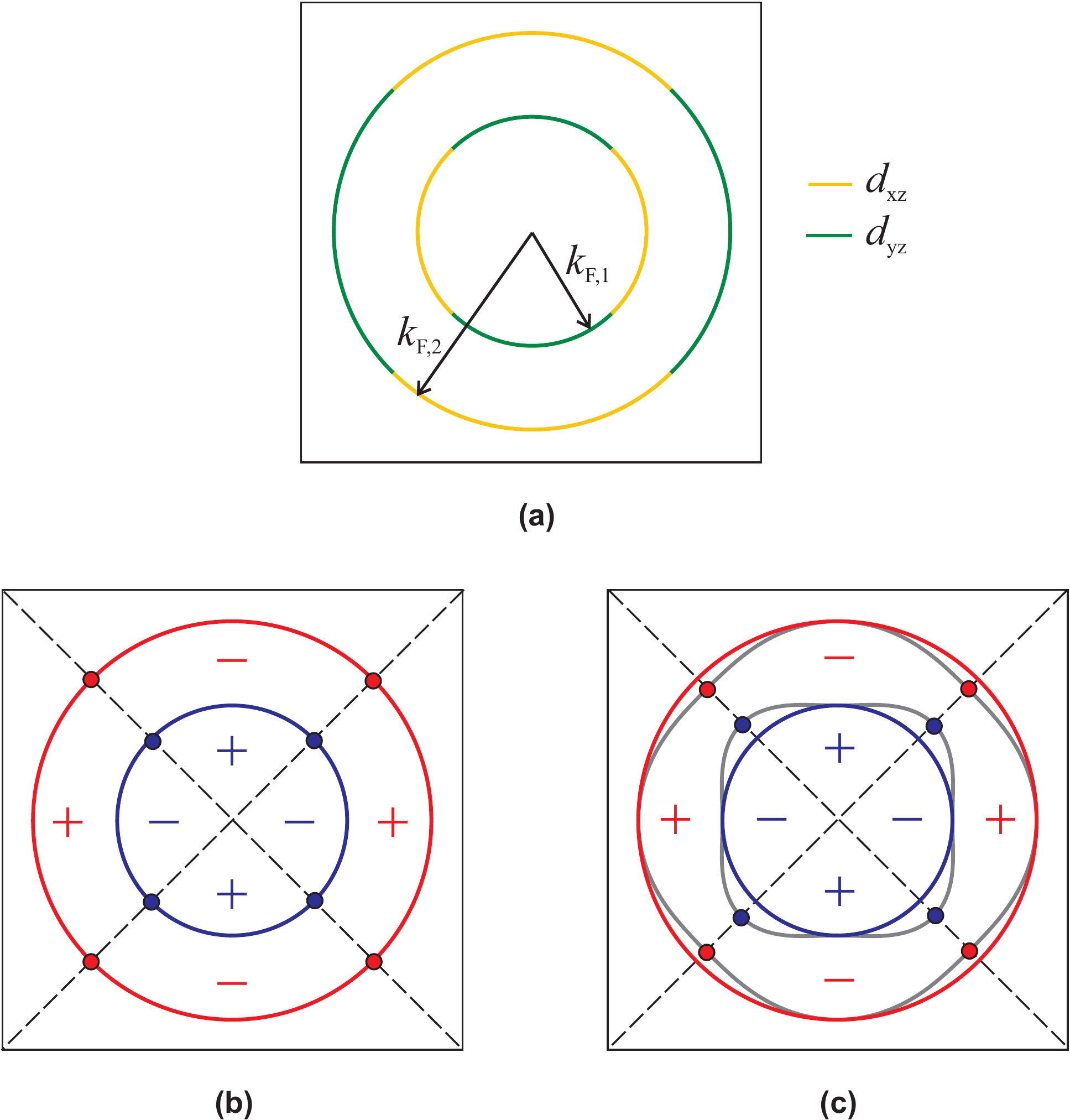}
\protect\protect\protect\protect\protect\protect\protect\protect\protect\caption{The Fermi surfaces (FS) and the location of the nodal points near
the two $\Gamma$-centered $d_{xz}/d_{yz}$ pockets. Panel (a) shows
the two FS in the normal state, highlighting the orbital that gives
the largest spectral weight at each point along the FS (yellow for
$d_{xz}$ and green for $d_{yz}$). Panel (b) illustrates the location
of the $d-$wave nodes on the two FS (blue and red lines) if the band
off-diagonal gap term was absent. Panel (c) presents the actual location
of the nodal points (red and blue dots) for the case $\Delta=0.8\Delta_{\mathrm{cr}}$.
The dispersions are given by $E_{a,b}=\sqrt{\Delta^{2}\cos^{2}{2\theta}+\epsilon_{a,b}^{2}}$
and the terms $\epsilon_{a,b}$ vanish along the two gray lines adjacent
to the FS. \label{fig:schematic} }
\end{figure}

The arguments in favor of a $d-$wave gap symmetry, however, have
been questioned by angle-resolved photoemission (ARPES) measurements~\cite{ARPES_KFEAS,shin}.
For hole-doped KFe$_{2}$As$_{2}$, these measurements have found~\cite{shin}
that the gap on the inner hole pocket centered at the $\Gamma$ point
(${\bf k}=0$) displays some angle variation but has no nodes~\cite{comm}.
The conventional wisdom is that a $d-$wave gap must vanish on all
Fermi surfaces (FSs) centered at ${\bf k}=0$ along symmetry-imposed
directions in momentum space -- specifically, a $d_{x^{2}-y^{2}}$
gap, which we consider hereafter, must vanish on the FS points along
the diagonals $k_{x}=\pm k_{y}$ in the 1-Fe Brillouin zone (1Fe BZ).
The non-vanishing of the gap on the inner FS along these direction
in ARPES measurements was interpreted~\cite{shin} as the smoking-gun
evidence ruling out a $d-$wave gap in KFe$_{2}$As$_{2}$. Similarly,
in K$_{x}$Fe$_{2-y}$Se$_{2}$, the gap has been measured on the
electron pocket centered at the $Z$-point ($k_{x}=k_{y}=0$ and $k_{z}=\pi$),
and was found to be almost angle-independent \cite{ARPES_Z}. Again,
the conventional wisdom is that this result is fundamentally inconsistent
with a $d-$wave gap symmetry.

It was argued in Ref. \cite{si} that the $d-$wave order parameter
in FeSCs necessarily contains both intra-pocket and inter-pocket components,
and by this reason a $d-$wave gap has no nodes along the Fermi surfaces.
A similar effect was previously shown to impact the behavior of accidental
nodes in an $s^{+-}$ superconductor \cite{alberto,alberto_1}. In
this paper, we revisit this issue and investigate the fate of the
$d-$wave nodes in FeSCs on the FSs centered at the high symmetry
$\Gamma$ and $Z$ points. We argue that one has to distinguish between
the cases when a FS centered at $k_{x}=k_{y}=0$ is made out of a
single orbital, like the $Z$-centered electron pocket of certain
compounds, and the cases when the FSs centered at these points are
made out of even number of orbitals, like the $\Gamma$-centered hole
pockets present in most compounds, which are made out of $d_{xz}/d_{yz}$
orbitals. In the first case, the symmetry-imposed $d-$wave nodes
remain on the FS. In the second case, the $d-$wave gap \textit{does
not} \emph{have nodes} on the normal state FSs (see Fig. \ref{fig:schematic}a.)
We demonstrate, however, that \emph{this does not imply that the electronic
spectrum is gapped}. We show that the nodes remain along the high-symmetry
directions, but get displaced from the original FSs, at least near
$T_{c}$, when the gaps are small. If the difference between the Fermi
momenta of the two pockets is substantial, the nodes persist down
to $T=0$. If, however, the pockets are close to each other, pairs
of nodes with opposite winding numbers can annihilate at $T_{\mathrm{cr}}<T_{c}$,
rendering the spectrum gapped.

The displacement of the nodes from the FSs is related to how intra-orbital
pairing in the orbital basis is displayed in the band basis\textbf{
}\cite{orbital_SC,si}. Namely, in the absence of spin-orbit interaction,
tetragonal symmetry requires that the $d-$wave gap on these pockets
must be diagonal in the orbital basis, i.e. $\left\langle d_{xz,-\mathbf{k}\downarrow}d_{xz,\mathbf{k}\uparrow}\right\rangle =\Delta,~\left\langle d_{yz,-\mathbf{k}\downarrow}d_{yz,\mathbf{k}\uparrow}\right\rangle =-\Delta$.
However, to analyze the gap structure near the FS, one needs to change
basis from orbital space to band space. The latter is characterized
by the band operators $c_{1,\mathbf{k}\sigma}$ and $c_{2,\mathbf{k}\sigma}$,
which describe excitations near the two hole FS. As a result, in band
basis, the same $d-$wave gap acquires both diagonal and off-diagonal
components: $\left\langle c_{1,-\mathbf{k}\downarrow}c_{1,\mathbf{k}\uparrow}\right\rangle =-\left\langle c_{2,-\mathbf{k}\downarrow}c_{2,\mathbf{k}\uparrow}\right\rangle =\Delta\cos2\theta$
and $\left\langle c_{2,-\mathbf{k}\downarrow}c_{1,\mathbf{k}\uparrow}\right\rangle =\left\langle c_{1,-\mathbf{k}\downarrow}c_{2,\mathbf{k}\uparrow}\right\rangle =\Delta\sin{2\theta}$,
respectively. For circular and small hole FS, $\theta$ coincides
with the angle along the FS.

Because the off-diagonal gap term mixes the two FS, the $d-$wave
gap varies as function of $\theta$ but does not have nodes. The strength
of the variation depends on the interplay between $\Delta$ and the
splitting between the two hole FS, as we discuss below. Such an effect
does not happen for an $s$-wave gap, since the orbital and band representations
are identical in this case, implying that off-diagonal terms do not
emerge, unless there is hybridization between the pockets. The orbital
and the band representations are also identical, for any gap symmetry,
on a pocket made out of a single orbital, such as the $d_{xy}$ $Z$-pocket
in K$_{x}$Fe$_{2-y}$Se$_{2}$.

\begin{figure}[h]
\centering{}\includegraphics[width=0.9\columnwidth]{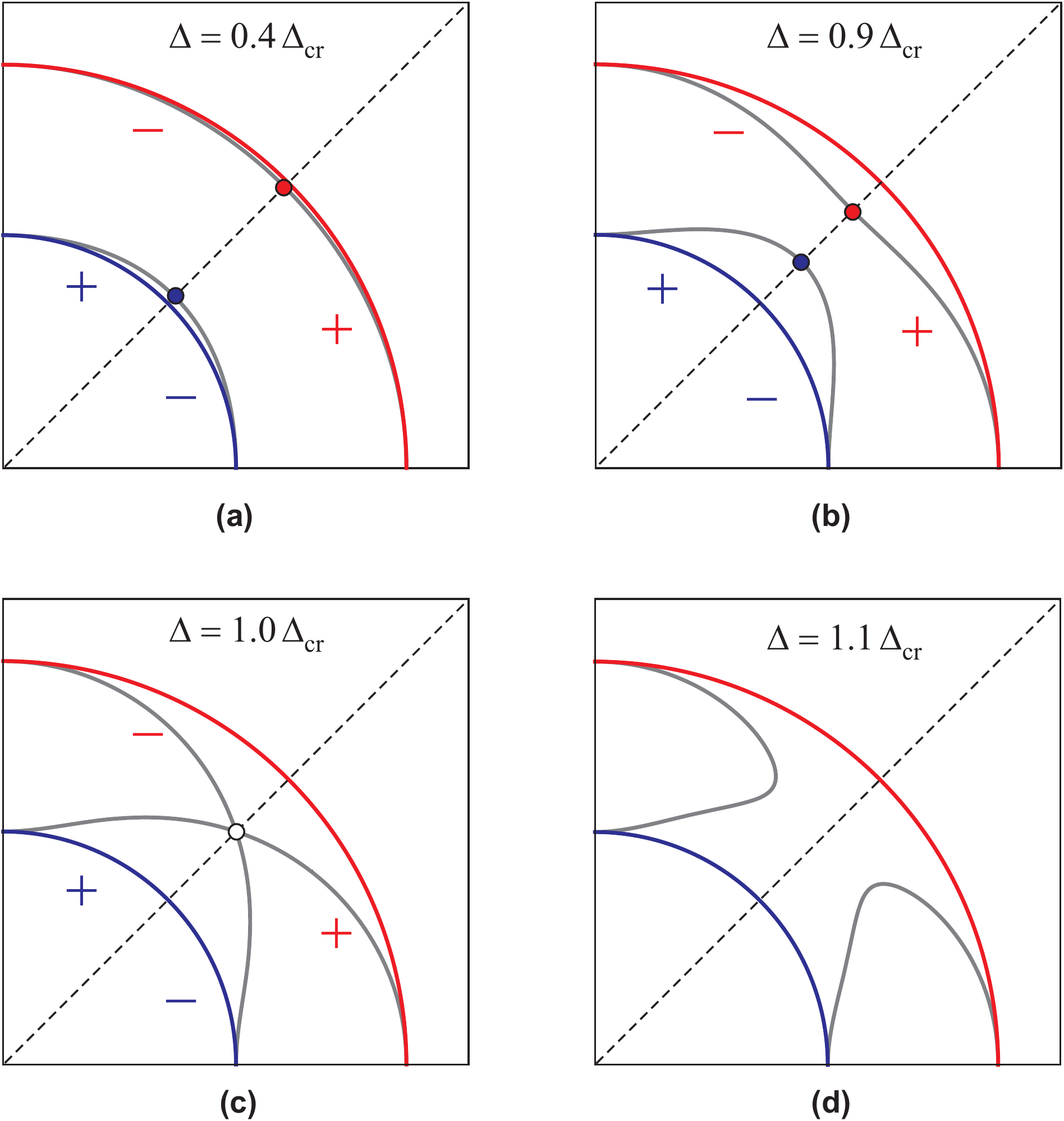}
\protect\protect\protect\protect\protect\protect\protect\protect\protect\caption{The evolution of the $d-$wave nodes as $\Delta$ increases beyond
the critical value $\Delta_{\mathrm{cr}}$. The blue and red lines
are the normal state FS. The gray lines denote the locations of $\epsilon_{a,b}=0$,
and the dispersions are given by $E_{a,b}=\sqrt{\Delta^{2}\cos^{2}{2\theta}+\epsilon_{a,b}^{2}}$.
The nodal points are marked by the red and blue dots. Four pairs of
nodal points are present for $\Delta<\Delta_{\mathrm{cr}}$ and disappear
for $\Delta>\Delta_{\mathrm{cr}}$. In this figure, we used circular
band dispersions with $m_{2}=3m_{1}$. \label{fig:nodes_evolution} }
\end{figure}

However, by extending the analysis to momenta away from the normal
state FS, we found that the nodes in the $d-$wave excitation spectrum
near the $d_{xz}/d_{yz}$ hole pockets do survive, and are just displaced
from the normal state hole FSs. Specifically, the excitation spectrum
has the form~\cite{si}
\begin{equation}
E_{a,b}^{2}=\Delta^{2}\cos^{2}{2\theta}+\epsilon_{a,b}^{2}\label{dispersions}
\end{equation}
with
\begin{align}
\epsilon_{a,b} & =\text{sgn}\left(\epsilon_{1,\mathbf{k}}+\epsilon_{2,\mathbf{k}}\right)\sqrt{\left(\frac{\epsilon_{1,\mathbf{k}}+\epsilon_{2,\mathbf{k}}}{2}\right)^{2}+\Delta^{2}\sin^{2}{2\theta}}\nonumber \\
 & \pm\left(\frac{\epsilon_{1,\mathbf{k}}-\epsilon_{2,\mathbf{k}}}{2}\right)\label{aux_dispersions}
\end{align}
where $\epsilon_{1,\mathbf{k}}$ and $\epsilon_{2,\mathbf{k}}$ are
the normal state dispersions of bands $1$ and $2$, respectively.
If the off-diagonal term $\Delta\sin{2\theta}$ was absent, $\epsilon_{a}=\epsilon_{1,\mathbf{k}}$,
$\epsilon_{b}=\epsilon_{2,\mathbf{k}}$, and the dispersions would
be the conventional ones for a $d-$wave SC, namely, $E_{a,b}^{2}=\Delta^{2}\cos^{2}{2\theta}+\epsilon_{1,2}^{2}$.
In this case, each dispersion would have nodal points on the FS at
$\theta=\theta_{n}\equiv\left(2n+1\right)\pi/4$, with $n=0,1,2,3$
(see Fig. \ref{fig:schematic}b). Because of the off-diagonal term,
however, $\epsilon_{a}$ does not vanish when $\epsilon_{1}=0$ and
$\epsilon_{b}$ does not vanish when $\epsilon_{2}=0$. However $\epsilon_{a}$
($\epsilon_{b}$) does vanish along the lines specified by $\epsilon_{1,\mathbf{k}}\epsilon_{2,\mathbf{k}}=-\Delta^{2}\sin^{2}{2\theta}$,
which are displaced from the actual FS, see Fig. \ref{fig:schematic}c.

When the magnitude of the $d-$wave gap is small, the two lines are
well separated and cross the direction $\theta=\theta_{n}$ at the
momenta $k_{a}>k_{F,1}$ and $k_{b}<k_{F,2}$. At these crossing points,
the full quasiparticle energy $E_{a}$ ($E_{b})$ vanishes. These
are new $d-$wave nodal points, shifted from their corresponding FS
by the mixing term. For small $\Delta$, this shift is small, of order
$\Delta^{2}$. However, as temperature decreases, $\Delta$ becomes
larger and the nodal points become closer. If the gap reaches the
critical value $\Delta_{\mathrm{cr}}$, which depends on the radii
of the two pockets, the two nodal points merge and annihilate each
other. In particular, at the Lifshitz transition taking place for
$\Delta=\Delta_{\mathrm{cr}}$, the $\epsilon_{a}=0$ and $\epsilon_{b}=0$
lines mix and split in the orthogonal direction, see Fig. \ref{fig:nodes_evolution}.
For $\Delta>\Delta_{\mathrm{cr}}$, these lines no longer cross the
directions $\theta=\theta_{n}$, i.e. $\epsilon_{a,b}$ and $\Delta\cos{2\theta}$
do not vanish simultaneously. In this situation, $E_{a,b}$ do not
have nodal points, implying that the excitations of the $d-$wave
superconductor are fully gapped.

When nodal points are present, the excitations near $E_{a,b}=0$ are
Dirac-cones, $E_{a,b}=\sqrt{{\tilde{k}}_{x}^{2}+{\tilde{k}}_{y}^{2}}$,
where $\tilde{x}$ and $\tilde{y}$ are directions along and transverse
to the lines $\epsilon_{a}=\epsilon_{b}=0$, defined by ${\tilde{k}}_{x}=2\Delta(\theta-\theta_{n})$
and ${\tilde{k}}_{y}=(\frac{d\epsilon_{a,b}}{dk})~(k-k_{a,b})$, where
the derivative is taken at $\theta=\theta_{n}$. At the critical gap
value $\Delta=\Delta_{\mathrm{cr}}$, because $d\epsilon_{a,b}/dk$
vanishes we find ${\tilde{k}}_{y}\propto(k-k_{a,b})^{2}$. This dispersion
has the same form as the dispersion of fermions at the critical point
between a semi-metal and an insulator~\cite{montambaux,me,moon}.
It was argued~\cite{me} that for such a dispersion the system with
dynamically screened Coulomb interaction should display a highly non-trivial
quantum-critical behavior in both fermionic and bosonic sectors. Our
study shows that a $d-$wave FeSC provides an interesting realization
of such behavior.

The displacement of the nodes to momenta away from the normal state
FSs has been previously discussed for accidental nodes on electron
pockets in an $s^{+-}$ superconductor. In this case, the displacement
is due to hybridization between these pockets~\cite{ku,alberto}.
The authors of Ref. \cite{alberto} argued that, as the hybridization
parameter gets larger, pairs of accidental nodes come close, and at
some critical hybridization merge and annihilate. The same effect
occurs~\cite{alberto_1} when one increases the pnictogen/chalcogen-induced
interaction between fermions on Fe sites (i.e. interaction with momentum
non-conservation by $(\pi,\pi)$ in the 1Fe BZ). For a $d-$wave superconductor,
the lifting of the nodes on the normal state FSs was first discussed
in Ref. \cite{si}. These authors concluded that, for arbitrary $\Delta$,
the nodes are lifted not only on the normal state FSs, but that the
whole electronic spectrum is generally gapped, except for possible
accidental nodes. We, on the contrary, argue that, at least for small
$\Delta$, the symmetry-imposed nodes do survive and just shift from
the original FSs to momenta located between the original FSs. This
is similar to what happens with the accidental nodes in an $s^{+-}$
superconductor in the presence of hybridization. From a generic perspective,
the persistence of the nodal points is associated with the fact that
each Dirac node has a non-zero winding number. Only when the two nodal
points with opposite winding numbers come close to each other under
the variation of some parameter (the magnitude of the gap in the $d-$wave
case), they can merge and annihilate. We discuss the comparison with
earlier works in more detail later in the paper.

We also emphasize that the nodal points in the $d-$wave case are
true symmetry-imposed $d-$wave nodes, and the damping near each nodal
point is the same as near a $d-$wave node on the FS in a conventional
case. Therefore, all thermodynamic properties of the system are also
the same as in a conventional $d-$wave superconductor. Only in ARPES
one can distinguish between a conventional $d$-wave case with nodes
on the original FS and the case when the nodes are shifted away from
the normal state FS due to the presence of the inter-pocket pairing
component.

The paper is organized as follows: in Section II, we introduce the
model, in Section III we derive the excitation spectrum, in Section
IV we compare our results with the case of a semi-metal to insulator
transition. In Sec. V we compare our results with earlier studies,
and in Section VI we present our conclusions.

\section{Model for $d$-wave superconductivity}

To focus on the main message of this paper, we consider a simplified
model of an FeSC with two $\Gamma$-centered hole pockets made out
of the $d_{xz}$ and $d_{yz}$ orbitals (Fig. \ref{fig:schematic}a),
and assume that 4-fermion interactions give rise to $d-$wave superconductivity
with $d_{x^{2}-y^{2}}$ gap symmetry (for a $d_{xy}$ gap symmetry,
the results are analogous to the ones that we obtain below). The attraction
in the $d-$wave channel may be due to the interactions within the
$d_{xz}/d_{yz}$ subset, as we assume for simplicity, or it can be
induced by the coupling to other orbitals. In the $d_{x^{2}-y^{2}}$
ordered state, which belongs to the $B_{1g}$ irreducible representation
of the $D_{4h}$ group, the gap function in the orbital basis is given
by $\left\langle d_{xz,-\mathbf{k}\downarrow}d_{xz,\mathbf{k}\uparrow}\right\rangle =\Delta,~\left\langle d_{yz,-\mathbf{k}\downarrow}d_{yz,\mathbf{k}\uparrow}\right\rangle =-\Delta$.
There are no inter-orbital terms $\left\langle d_{yz,-\mathbf{k}\downarrow}d_{xz,\mathbf{k}\uparrow}\pm d_{xz,-\mathbf{k}\downarrow}d_{yz,\mathbf{k}\uparrow}\right\rangle $
as they belong to the different irreducible representations $B_{2g}$
(plus sign) and $A_{2g}$ (minus sign).

Although the anomalous terms $\left\langle d_{i,-\mathbf{k}\downarrow}d_{j,\mathbf{k}\uparrow}\right\rangle $
are diagonal, the kinetic energy near the $\Gamma$ point does contain
terms describing hoping from one orbital to the other. The kinetic
energy is diagonalized by converting from the orbital to the band
basis, yielding:
\begin{equation}
{\cal H}_{0}=\sum_{\mathbf{k},\alpha}\left(\epsilon_{1,\mathbf{k}}c_{1,\mathbf{k}\alpha}^{\dagger}c_{1,\mathbf{k}\alpha}+\epsilon_{2,\mathbf{k}}c_{2,\mathbf{k}\alpha}^{\dagger}c_{2,\mathbf{k}\alpha}\right).\label{a_2}
\end{equation}

The dispersions $\epsilon_{1,k}$ and $\epsilon_{2,k}$ are $C_{4}$-symmetric.
We assume for simplicity that the system parameters are such that
the hole pockets can be approximated as circular \cite{Chubukov_16},
i.e. $\epsilon_{1,\mathbf{k}}=\mu-k^{2}/(2m_{1})$ and $\epsilon_{2,\mathbf{k}}=\mu-k^{2}/(2m_{2})$.
The two dispersions are not identical when $m_{1}\neq m_{2}$, but
are degenerate by symmetry at $\mathbf{k}=0$ in the absence of spin-orbit
coupling~\cite{Cvetkovic2013,rafael_oskar}.

\begin{figure}[h]
\centering{}\includegraphics[width=0.9\columnwidth]{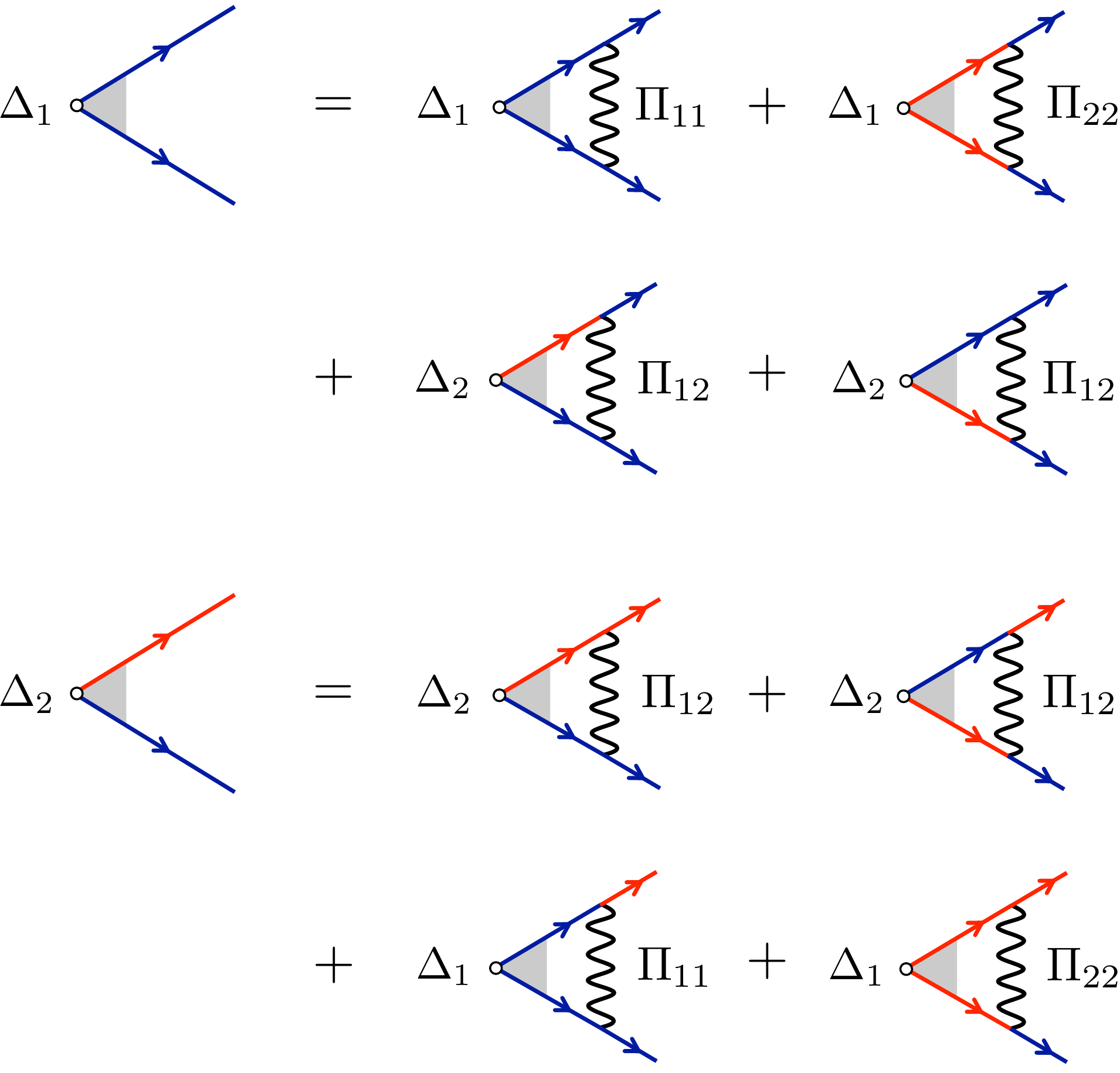} \protect\protect\protect\protect\protect\protect\protect\protect\protect\caption{The diagrammatic representation of the linearized gap equations, Eqs.
\protect\ref{a_7}. Blue and red lines denote fermions from bands
$c_{1}$ and $c_{2}$. \label{fig:4} }
\end{figure}

The transformation from the orbital operators $d_{xz}/d_{yz}$ to
the band operators $c_{1}$ and $c_{2}$ is a $U(1)$ rotation:
\begin{eqnarray}
 &  & d_{xz,\mathbf{k}\alpha}=\cos{\theta}_{\mathbf{k}}c_{1,\mathbf{k}\alpha}+{\sin\theta}_{\mathbf{k}}c_{2,\mathbf{k}\alpha},\nonumber \\
 &  & d_{yz,\mathbf{k}\alpha}=\cos{\theta}_{\mathbf{k}}c_{2,\mathbf{k}\alpha}-\sin{\theta}_{\mathbf{k}}c_{1,\mathbf{k}\alpha},~~\label{a_1}
\end{eqnarray}

For circular Fermi pockets the rotation angle $\theta_{\mathbf{k}}$
coincides with the polar angle $\theta$ along the FS \cite{Chubukov_16}.
Using Eq. (\ref{a_1}) we also re-express the anomalous term ${\cal H}_{\Delta}=\Delta\sum_{\mathbf{k}}\left(d_{xz,\mathbf{k}\uparrow}^{\dagger}d_{xz,-\mathbf{k}\downarrow}^{\dagger}-d_{yz,\mathbf{k}\uparrow}^{\dagger}d_{yz,-\mathbf{k}\downarrow}^{\dagger}\right)$
in the band basis. We obtain a combination of inter-band and intra-band
terms:

\begin{align}
 & {\cal H}_{\Delta}=\Delta_{a}\sum_{\mathbf{k}}\left(i\sigma_{\alpha\beta}^{y}\right)\left(c_{1,\mathbf{k}\alpha}^{\dagger}c_{1,-\mathbf{k}\beta}^{\dagger}-c_{2,\mathbf{k}\alpha}^{\dagger}c_{2,-\mathbf{k}\beta}^{\dagger}\right)+\nonumber \\
 & \Delta_{b}\sum_{\mathbf{k}}\left(i\sigma_{\alpha\beta}^{y}\right)\left(c_{1,\mathbf{k}\alpha}^{\dagger}c_{2,-\mathbf{k}\beta}^{\dagger}+c_{2,\mathbf{k}\alpha}^{\dagger}c_{1,-\mathbf{k}\beta}^{\dagger}\right)+h.c\label{a_3}
\end{align}
where $\sigma$ are Pauli matrices and in the $d-$wave case $\Delta_{a}=\Delta\cos2\theta$
and $\Delta_{b}=\Delta\sin2\theta$. Without loss of degeneracy, one
can set $\Delta_{a}$ to be real. $\Delta_{b}$ is, in general, a
complex variable.

Note that in the $d-$wave case, the inter-band anomalous terms are
of the same order $\Delta$ as intra-band terms and differ only by
their angular dependence. This may seem counterintuitive, as the pairing
kernel involving fermions from different bands is much smaller than
the kernel involving fermions from the same band. To see why inter-band
and intra-band pairing terms are nevertheless comparable, one can
explicitly solve for the intra-band and inter-band pairing vertices
by using a microscopic interaction that favors $d-$wave. For concreteness,
consider a toy model with pair-hopping interaction:
\begin{eqnarray}
H_{\mathrm{int}}=\frac{g}{2}\sum\left[d_{xz\alpha}^{\dag}d_{yz\alpha}d_{xz\beta}^{\dag}d_{yz\beta}+d_{yz\alpha}^{\dag}d_{xz\alpha}d_{yz\beta}^{\dag}d_{xz\beta}\right]\label{a_4}
\end{eqnarray}
where the summation over momenta and spin indices is left implicit.
A positive $g$ favors $B_{1g}$ pairing as one can verify in a straightforward
way by solving the gap equation in the orbital basis. Converting this
Hamiltonian into band basis and projecting onto the $B_{1g}$ channel,
we obtain \begin{widetext}
\begin{equation}
H_{\mathrm{int}}=-\frac{g}{4}\sum\left[\eta_{1,\mathbf{k}}^{\dagger}\eta_{1,\mathbf{p}}\cos{2\theta_{\mathbf{k}}}\cos{2\theta_{\mathbf{p}}}+\eta_{2,\mathbf{k}}^{\dagger}\eta_{2,\mathbf{p}}\sin{2\theta_{\mathbf{k}}}\sin{2\theta_{\mathbf{p}}}+\left(\eta_{1,\mathbf{k}}^{\dagger}\eta_{2,\mathbf{p}}+\eta_{2,\mathbf{k}}^{\dagger}\eta_{1,\mathbf{p}}\right)\sin{2\theta_{\mathbf{p}}}\cos{2\theta_{\mathbf{k}}}\right]\label{a_6}
\end{equation}
\end{widetext} where $\eta_{1,\mathbf{k}}^{\dagger}=c_{1,\mathbf{k}\alpha}^{\dagger}c_{1,-\mathbf{k}\beta}^{\dagger}-c_{2,\mathbf{k}\alpha}^{\dagger}c_{2,-\mathbf{k}\beta}^{\dagger}$,
$\eta_{2,\mathbf{k}}^{\dagger}=c_{1,\mathbf{k}\alpha}^{\dagger}c_{2,-\mathbf{k}\beta}^{\dagger}+c_{2,\mathbf{k}\alpha}^{\dagger}c_{1,-\mathbf{k}\beta}^{\dagger}$,
and the summation is over momentum and spin indices. Introducing the
two anomalous vertices $\Delta_{1}\left(i\sigma_{\alpha\beta}^{y}\right)\eta_{1,\mathbf{k}}^{\dagger}\cos2\theta_{\mathbf{k}}$
and $\Delta_{2}\left(i\sigma_{\alpha\beta}^{y}\right)\eta_{2,\mathbf{k}}^{\dagger}\sin2\theta_{\mathbf{k}},$
and solving the BCS-like gap equations shown graphically in Fig. \ref{fig:4},
we obtain
\begin{eqnarray}
 &  & \Delta_{1}=\frac{g}{4}\left(\frac{\Pi_{11}+\Pi_{22}}{2}\right)\Delta_{1}+\frac{g}{4}\Pi_{12}\Delta_{2}\nonumber \\
 &  & \Delta_{2}=\frac{g}{4}\Pi_{12}\Delta_{2}+\frac{g}{4}\left(\frac{\Pi_{11}+\Pi_{22}}{2}\right)\Delta_{1}\label{a_7}
\end{eqnarray}
where $\Pi_{11},\Pi_{22}$, and $\Pi_{12}$ (all positive) are particle-particle
polarization bubbles made out of $c_{1}$ and $c_{2}$ fermions in
the superconducting state. Near $T_{c}$, we obtain
\begin{eqnarray}
 &  & \Pi_{11}=\frac{1}{2}\int d^{2}\mathbf{k}\,\frac{\tanh\left(\frac{\epsilon_{1,\mathbf{k}}}{2T}\right)}{\left|\epsilon_{1,\mathbf{k}}\right|},~\Pi_{22}=\frac{1}{2}\int d^{2}\mathbf{k}\,\frac{\tanh\left(\frac{\epsilon_{2,k}}{2T}\right)}{\left|\epsilon_{2,\mathbf{k}}\right|},\nonumber \\
 &  & \Pi_{12}=\frac{1}{2}\int d^{2}\mathbf{k}\,\frac{\tanh\left(\frac{\epsilon_{1,\mathbf{k}}}{2T}\right)+\tanh\left(\frac{\epsilon_{2,\mathbf{k}}}{2T}\right)}{\left|\epsilon_{1,\mathbf{k}}+\epsilon_{2,\mathbf{k}}\right|}.\label{a_7_1}
\end{eqnarray}

Comparing the two expressions in Eq. (\ref{a_7}), we see that $\Delta_{1}=\Delta_{2}=\Delta$,
no matter what is the ratio of the inter-pocket and intra-pocket polarization
operators. This holds as long as the interaction $g$ is momentum-independent.
If momentum dependence is included, the intra-pocket and inter-pocket
interaction terms in (\ref{a_6}) differ more than by their distinct
angular dependences. In this situation, the r.h.s. of the two equations
in (\ref{a_7}) are no longer identical, and generally $\Delta_{1}>\Delta_{2}$.
In the limiting case $\Delta_{2}\to0$ one recovers the conventional
case with only intra-band pairing condensate.

\begin{figure}[h]
\centering{}\includegraphics[width=0.9\columnwidth]{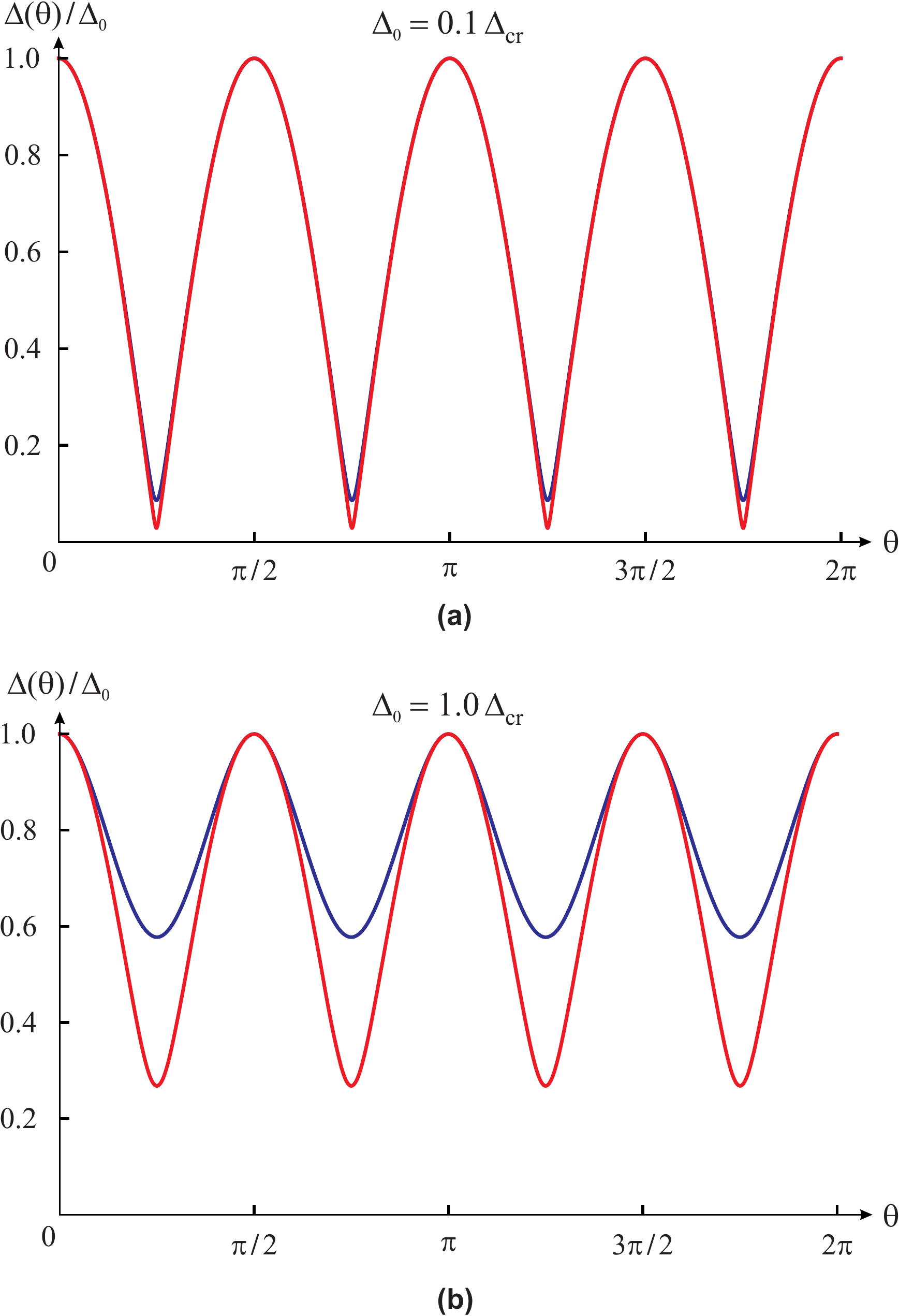}
\protect\protect\protect\protect\protect\protect\protect\protect\protect\caption{The dispersion of the $d-$wave gap along the two FS, Eq. \protect\ref{a_11},
for two values of $\Delta$. Blue (red) lines denote band $c_{1}$
($c_{2}$). There is substantial angular variation but no nodes. Note
also that the minimum value of the gap is different in both bands.
\label{fig:dispersion} }
\end{figure}

\section{Excitation spectrum}

We now return to Eqs. (\ref{a_2}) and (\ref{a_3}). The quadratic
Hamiltonian ${\cal H}_{0}+{\cal H}_{\Delta}$ can be straightforwardly
diagonalized and yields
\begin{equation}
{\cal H}=\sum_{\mathbf{k},\alpha}E_{a}(\mathbf{k})a_{\mathbf{k}\alpha}^{\dag}a_{\mathbf{k}\alpha}+\sum_{\mathbf{k},\alpha}E_{b}(\mathbf{k})b_{\mathbf{k}\alpha}^{\dag}b_{\mathbf{k}\alpha}\label{a_8}
\end{equation}
where
\begin{eqnarray}
 &  & E_{a,b}^{2}(\mathbf{k})=\frac{\epsilon_{1,\mathbf{k}}^{2}+\epsilon_{2,\mathbf{k}}^{2}}{2}+\Delta_{a}^{2}+|\Delta_{b}^{2}|\pm\nonumber \\
 &  & \sqrt{\left(\frac{\epsilon_{1,\mathbf{k}}^{2}-\epsilon_{2,\mathbf{k}}^{2}}{2}\right)^{2}+(\epsilon_{1,\mathbf{k}}-\epsilon_{2,\mathbf{k}})^{2}|\Delta_{b}|^{2}+4\Delta_{a}^{2}(\mathrm{Re}\Delta_{b})^{2}}\label{a_99}
\end{eqnarray}

In the $d-$wave case ($\Delta_{a}=\Delta\cos2\theta$ and $\Delta_{b}=\Delta\sin2\theta$),
Eq. (\ref{a_99}) can be simplified to
\begin{eqnarray}
 &  & E_{a,b}(\mathbf{k})=\sqrt{\Delta^{2}\cos^{2}{2\theta}+\epsilon_{a,b}^{2}(\mathbf{k})},\label{a_9}
\end{eqnarray}
where
\begin{align}
\epsilon_{a,b}(\mathbf{k}) & =\text{sgn}\left(\epsilon_{1,\mathbf{k}}+\epsilon_{2,\mathbf{k}}\right)\sqrt{\left(\frac{\epsilon_{1,\mathbf{k}}+\epsilon_{2,\mathbf{k}}}{2}\right)^{2}+\Delta^{2}\sin^{2}{2\theta}}\nonumber \\
 & \pm\left(\frac{\epsilon_{1,\mathbf{k}}-\epsilon_{2,\mathbf{k}}}{2}\right)\label{a_10}
\end{align}

Eq. (\ref{a_99}) was first obtained in Ref. \cite{alberto} for an
$s^{+-}$ superconductor with accidental nodes ($\Delta_{a}=\Delta$,
$\Delta_{b}=i\Delta\alpha\cos{2\theta}$, $\alpha>1$). For a $d-$wave
superconductor, Eqs. (\ref{a_9}), (\ref{a_10}) were first derived
in Ref. \cite{si}.

The dispersions $E_{a,b}$ in Eqs. (\ref{a_9}), (\ref{a_10}) have
the same forms as in a conventional $d$-wave superconductor, but
are actually more complex because $\epsilon_{a,b}$ themselves depend
on $\Delta$. For a vanishing $\Delta$, $\epsilon_{a}$ and $\epsilon_{b}$
coincide with the normal state dispersions, $\epsilon_{a}=\epsilon_{1,\mathbf{k}}$
and $\epsilon_{b}=\epsilon_{2,\mathbf{k}}$, as they indeed should.
At a finite but small $\Delta$ (i.e., near $T_{c}$), $\epsilon_{a}=\epsilon_{1,\mathbf{k}}+\frac{\Delta^{2}\sin^{2}{2\theta}}{\epsilon_{1,\mathbf{k}}+\epsilon_{2,\mathbf{k}}}$
and $\epsilon_{b}=\epsilon_{2,\mathbf{k}}+\frac{\Delta^{2}\sin^{2}{2\theta}}{\epsilon_{1,\mathbf{k}}+\epsilon_{2,\mathbf{k}}}$.
We see that $\epsilon_{a}$ ($\epsilon_{b}$) \textit{does not} vanish
on the FS, where $\epsilon_{1}=0$ ($\epsilon_{2}=0$), except along
the particular directions $\sin{2\theta}=0$. For such values of $\theta$,
however, $\Delta^{2}\cos2\theta$ has a maximum value $\Delta^{2}$.
As a result, there are no zeroes of $E_{a,b}$ along each of the two
FSs, despite the fact that the gap is $d-$wave. At arbitrary $T<T_{c}$
we have at $\epsilon_{1,\mathbf{k}}=0$ (and $\epsilon_{2,\mathbf{k}}>0$)
\begin{equation}
E_{a}=\sqrt{\Delta^{2}\cos^{2}{2\theta}+\left(\sqrt{\frac{\epsilon_{2,k}^{2}}{4}+\Delta^{2}\sin^{2}{2\theta}}-\frac{\epsilon_{2,k}}{2}\right)^{2}}\label{a_11}
\end{equation}

We plot the excitation energies $E_{a}$ and $E_{b}$ as a function
of $\theta$ along both FS in Fig. \ref{fig:dispersion} for two values
of $\Delta$. We see that there is substantial angular variation of
$E_{a,b}(\theta)$, but no nodes.

We now analyze the excitation energies $E_{a,b}$ away from the FS.
A straightforward analysis of Eq. (\ref{a_10}) shows that $\epsilon_{a,b}$
vanish along the lines where
\begin{equation}
\epsilon_{1,\mathbf{k}}~\epsilon_{2,\mathbf{k}}=-\Delta^{2}\sin^{2}{2\theta}\label{a_12}
\end{equation}

For small $\Delta$ (i.e, near $T_{c}$), Eq. (\ref{a_12}) is satisfied
along two separate lines, one adjacent to the inner FS, ($\epsilon_{1,\mathbf{k}}=0$)
and another adjacent to the outer FS ($\epsilon_{2,\mathbf{k}}=0$).
We show the lines $\epsilon_{a}=0$ and $\epsilon_{b}=0$ in Fig \ref{fig:nodes_evolution}
for different values of $\Delta$. Because these lines cross the directions
along which $\cos{2\theta}=0$, $E_{a}$ or $E_{b}$ vanish at the
crossing points, i.e. \textit{the full excitation energy vanishes}.
This implies that the nodal points of the $d-$wave superconductor
still exist near $T_{c}$, but get shifted away from the normal state
FS by the inter-band component of the $d-$wave gap. The nodal points
are located along $\cos{2\theta}=0$, at $k=k_{a,b}$ given by
\begin{equation}
k_{a,b}^{2}=\left(\frac{k_{F,1}^{2}+k_{F,2}^{2}}{2}\right)\pm\sqrt{\left(\frac{k_{F,1}^{2}-k_{F,2}^{2}}{2}\right)^{2}-4m_{1}m_{2}\Delta^{2}},\label{a_14}
\end{equation}
where $k_{F,i}^{2}=2m_{i}\mu$.

The behavior of $E_{a,b}$ at smaller temperatures depends on the
interplay between the gap value and the difference between $m_{2}$
and $m_{1}$, or specifically, between $\Delta(T)$ and
\begin{equation}
\Delta_{\mathrm{cr}}=\mu\left(\frac{m_{2}-m_{1}}{2\sqrt{m_{1}m_{2}}}\right)\label{a_15}
\end{equation}

If $\Delta_{\mathrm{cr}}$ is large enough, the nodes survive down
to $T=0$. However, if $m_{2}-m_{1}$ is small enough (i.e., the inner
and the outer hole pockets are close), $\Delta(T)$ reaches $\Delta_{\mathrm{cr}}$
at some $T=T_{\mathrm{cr}}$ below $T_{c}$. At this temperature,
a Lifshitz transition occurs when the two nodal points merge at $k=k_{\mathrm{cr}}$
and then split in orthogonal directions, see Fig. \ref{fig:nodes_evolution}c.

On a technical side, we found that, when $\Delta$ is slightly below
$\Delta_{\mathrm{cr}}$, the two nodal points of the dispersion are
the nodes of $\epsilon_{b}$ (and $E_{b}$), while the dispersion
$\epsilon_{a}$ has no nodes. The change of the behavior from the
nodes in both $\epsilon_{a}$ and $\epsilon_{b}$ to two nodes in
$\epsilon_{b}$ occurs when $\Delta$ reaches the value $\Delta^{*}=\mu\left(\frac{m_{2}-m_{1}}{m_{1}+m_{2}}\right)$,
which is comparable but smaller than $\Delta_{\mathrm{cr}}$. The
ratio $\Delta^{*}/\Delta_{\mathrm{cr}}=2\sqrt{m_{1}m_{2}}/\left(m_{1}+m_{2}\right)<1$.
This change does not affect the location of the zeros of $\epsilon_{a,b}$
in momentum space (gray lines in Fig. Fig. \ref{fig:nodes_evolution}),
just the identification of these lines with $\epsilon_{a}$ or $\epsilon_{b}$
becomes more complex.

At $\Delta>\Delta_{\mathrm{cr}}$, the lines where $\epsilon_{a,b}=0$
form four disconnected closed loops (see Fig. \ref{fig:nodes_evolution}d).
Along these loops the excitation energy becomes $E=\Delta|\cos{2\theta}|$.
However, because the closed loops do not cross the directions $\cos{2\theta}=0$,
the nodes disappear, i.e. the excitation spectrum of a $d-$wave superconductor
becomes \textit{fully gapped}.

\section{Analogy with semi-metal to insulator transition}

There is a close analogy between the Lifshitz transition at $T=T_{\mathrm{cr}}$
in our problem and the transition from a 2D massless Dirac semi-metal
to an insulator. In the latter case, the semi-metal phase has two
separate Dirac nodal points with the winding numbers $\pm1$ \cite{oskar_1}.
Upon variation of some system parameter (e.g., strain in graphene),
the distance between the two nodal points decreases until they merge
and annihilate at a critical value of such parameter. At the critical
point, the system is described by fermions with linear dispersion
in one spatial direction and quadratic in the other.~\cite{montambaux,me,moon}
Similarly, in our case, near $T_{c}$, the dispersion along one of
the four directions specified by $\cos{2\theta}=0$ has two nodal
points with Dirac-like dispersions. Just as in the semi-metal to insulator
transition, the winding numbers near the two Dirac points are $\pm1$.
At $T=T_{\mathrm{cr}}$ (if it exists), the Dirac points merge. At
this temperature the excitation spectrum around the single nodal point
is quadratic along the direction in which $\cos{2\theta}=0$ and linear
in the transverse direction. At a smaller temperature, the excitation
spectrum is fully gapped, like in an insulator. Recent studies of
the semi-metal to insulator transition have shown~\cite{me} that
at the critical point the dynamically screened Coulomb interaction
gives rise to a highly non-trivial quantum-critical behavior in both
fermionic and bosonic sectors. A $d-$wave state in the FeSC will
provide a realization of such behavior if $T_{\mathrm{cr}}$ can be
tuned to zero by changing some external parameter, such as pressure.
An $s^{+-}$ superconductor, in which accidental nodes can be lifted
by varying an external parameter~\cite{alberto,alberto_1}, is another
realization of such semi-metal to insulator transition \cite{quadratic_nodes1,quadratic_nodes2,quadratic_nodes3}.

\section{Comparison with earlier works}

\begin{figure}[h]
\begin{centering}
\includegraphics[width=0.95\columnwidth]{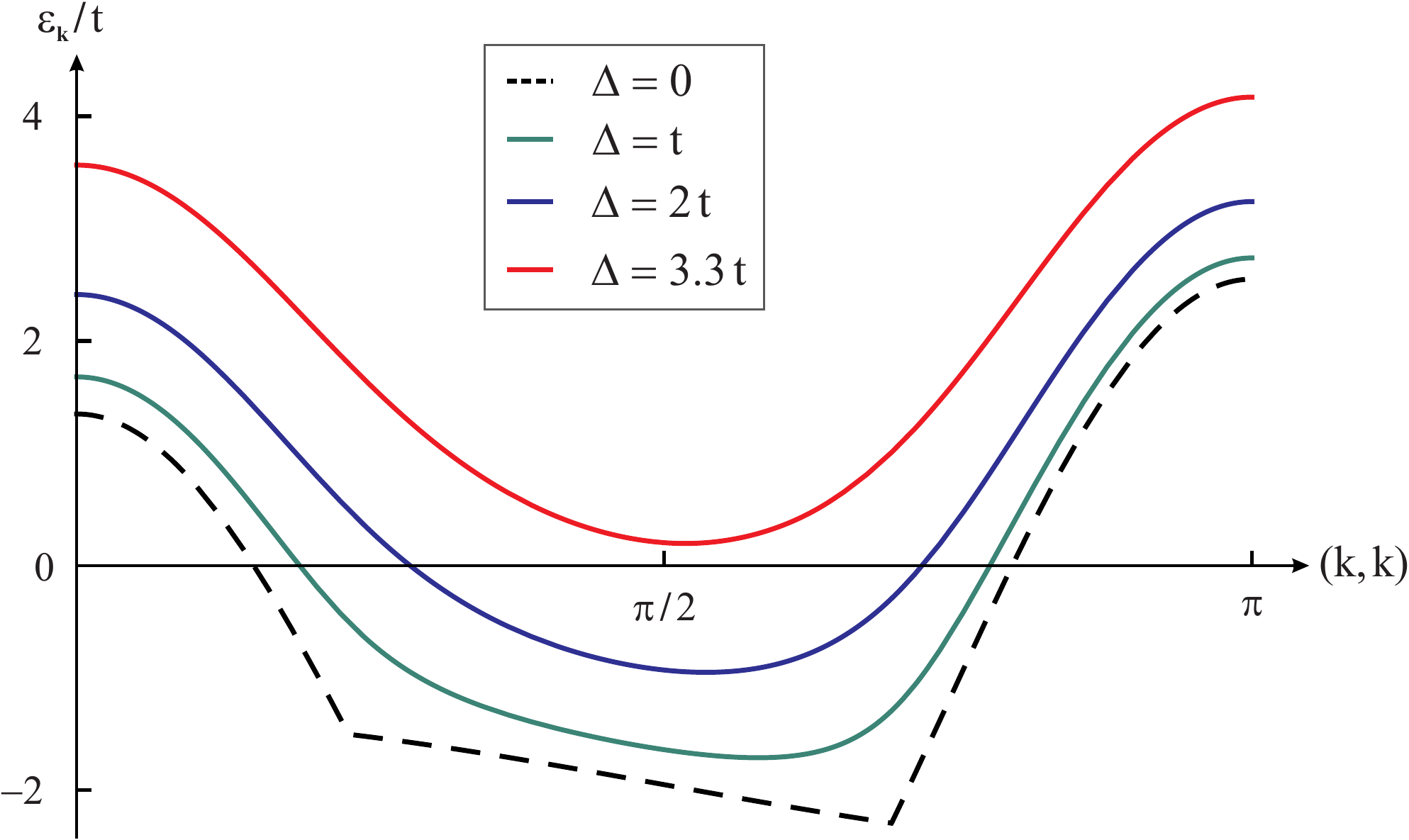}
\par\end{centering}

\protect\protect\protect\protect\caption{The quasiparticle dispersion in the $d-$wave superconducting state
for the two-orbital model of Raghu \emph{et al}. \cite{raghu}. The
dispersion has the form $E_{\mathbf{k}}^{2}=\epsilon_{\mathbf{k}}^{2}+\Delta^{2}\cos^{2}{2\theta_{\mathbf{k}}}$,
where $\epsilon_{\mathbf{k}}=(\xi_{+}^{2}+\Delta^{2}\sin^{2}2\theta_{\mathbf{k}})^{1/2}\pm|\vec{B}|$
(see Eqs. (\protect\ref{a_9}), (\protect\ref{a_10}) and Ref. \protect\cite{si}).
Ref. \protect\cite{si} used the parameters from Ref. \protect\cite{raghu}:
$\xi_{+}=-0.3t(\cos{k_{x}}+\cos{k_{y}})+3.4t\cos{k_{x}}\cos{k_{y}}-1.45t$,
$|\vec{B}|=t\left[2.3^{2}(\cos{k_{x}}-\cos k_{y})^{2}+3.4^{2}\sin^{2}{k_{x}}\sin^{2}{k_{y}}\right]^{1/2}$,
and $\sin{2\theta_{\mathbf{k}}}=3.4t\sin{k_{x}}\sin{k_{y}}/|\vec{B}|$,
where $t$ sets the overall energy scale. Nodal points of the dispersion
are the ones for which $\cos{2\theta_{\mathbf{k}}}=0$ and $\epsilon_{\mathbf{k}}=0$.
We plot $\epsilon_{\mathbf{k}}$ as a function of momentum ${\bf k}$
along the direction $k=k_{x}=k_{y}$ (for which $\cos{2\theta_{\mathbf{k}}}=0$)
for different gap values $\Delta$. A pair of nodal points is observed
unless $\Delta$ exceeds the critical value $\Delta_{c}\approx3.08t$,
which is about a quarter of the bandwidth. \label{fig:comparison} }
\end{figure}

Several earlier studies of the pairing involving fermions from two
different bands have already pointed out that an intra-pocket pairing
condensate generates an inter-band pairing condensate, generally of
the same order as the intra-band one \cite{khodas,si}. Ref. \cite{khodas}
focused on the system with only electron pockets. When inter-pocket
repulsion is dominant, the analysis within 1Fe BZ shows that the system
develops $d$-wave superconductivity with sign change of the gap on
the two electron pockets~\cite{peter}. However, the result holds
only as long as one neglects the coupling between electron pockets,
i.e., the processes with momentum non-conservation by $(\pi,\pi)$
in the 1Fe BZ. Hybridization, which is the combined effect of glide
plane symmetry and of spin-orbit interaction, triggers the appearance
of a inter-pocket pairing condensate in terms of the band fermions,
in analogy to what happens in our case. This effect does not affect
substantially the $d-$wave gap on the electron pockets, which in
2D has no nodes anyway, but it gives rise to a novel $s^{+-}$ pairing
between inner and outer hybridized electron pockets~\cite{khodas,mazin_1},
when the coupling associated with the hybridization exceeds a certain
critical value.

A shift of the nodal points to momenta away from the FS and their
subsequent merging and annihilation (a Lifshitz transition) has been
analyzed in several publications \cite{ku,alberto,alberto_1} in the
context of the behavior of accidental nodes under a change of system
parameters such as hybridization~\cite{alberto}, interaction with
momentum non-conservation by $(\pi,\pi)$ in the 1Fe BZ~\cite{alberto_1},
or application of strain \cite{Kang14}. The key features in the $s^{+-}$
case are the same as in the $d-$wave case, namely, under a change
of some parameter, which induces inter-pocket pairing term in the
band basis, nodal points initially survive, but shift away from the
FS, into the region between the pockets (electron pockets in $s^{+-}$
case). As the strength of the inter-pocket pairing term increases,
neighboring nodal points come closer to each other and eventually
merge and annihilate. There is one distinction to our case, however
-- the merging of accidental nodes in $s-$wave superconductor involves
neighboring nodal points which were originally on the same FS, i.e.,
nodal points have to travel in the direction along the FS.

The non-trivial interplay between the $d-$wave order parameter in
the orbital and the band basis has been first analyzed in Ref. \cite{si}.
The authors of~\cite{si} correctly pointed out that the inter-band
pairing component makes the excitations along the FS nodeless despite
that the gap has a d-wave symmetry. In Ref. \cite{si} the $d-$wave
order parameter in the orbital basis was assumed to have the form
$\Delta(\mathbf{k})=g_{\mathbf{k}}\left\langle d_{xz,\mathbf{k}\uparrow}^{\dagger}d_{xz,-\mathbf{k}\downarrow}^{\dagger}-d_{yz,\mathbf{k}\uparrow}^{\dagger}d_{yz,-\mathbf{k}\downarrow}^{\dagger}\right\rangle $,
with $g_{\mathbf{k}}$ changing sign between hole and electron pockets
(such an order parameter has been listed previously among other singlet
pairing order parameters in Eq. D1 of Ref. \cite{Cvetkovic2013}).
For the purposes of comparison with our paper, where only hole pockets
are studied, it is sufficient to consider $g_{\mathbf{k}}$ near hole
pockets, where it can be approximated by a constant.

Our result for the electronic dispersion, Eqs. (\ref{a_9}) and (\ref{a_10}),
reproduces Eq. (5) of Ref. \cite{si}, yet the conclusions are somewhat
different. The authors of Ref.\cite{si} concluded that the presence
of inter-pocket pairing component makes the electronic spectrum generically
gapped, except for possible accidental nodes. We, on the contrary,
argue that the true symmetry-imposed $d-$wave nodal points survive,
at least near $T_{c}$, and just shift away from the normal state
FS. We further argue that a $d-$wave superconductor can be fully
nodeless, but this happens only when pairs of nodal points with opposite
winding numbers come close, merge, and annihilate. The presence of
two nearly-located hole FSs is crucial for this last effect, otherwise
the critical $\Delta_{\mathrm{cr}}$, above which the spectrum becomes
nodeless, is comparable to the bandwidth, and the gap $\Delta$ necessary
remains smaller than $\Delta_{\mathrm{cr}}$ down to $T=0$.

As one illustration of their analysis, the authors of Ref. \cite{si}
considered the two-orbital lattice model with tight-binding parametrization
of Ref. \cite{raghu}. This model is different from two-orbital low-energy
model and has one hole pocket at the center of the 1Fe BZ and another
hole pocket at the corner of the 1Fe BZ. We argue that in this model,
$\Delta_{\mathrm{cr}}$ is large -- a fraction of the bandwidth. To
demonstrate this, in Fig. \ref{fig:comparison} we plot the dispersion
along the $k_{x}=k_{y}$ direction, showing that the symmetry-imposed
$d-$wave nodes are indeed present at $\Delta$ smaller than the hopping
integral $t$, only their position shifts from the normal state FS.
The nodes annihilate and fully gapless spectrum appears only for $\Delta>\Delta_{\mathrm{cr}}=3.08t$.
The large value of $\Delta_{\mathrm{cr}}$ is due to the fact that
the two hole pockets are centered at different points of the 1Fe BZ.
When both hole pockets are centered at $\Gamma$, $\Delta_{\mathrm{cr}}$
is much smaller.

We emphasize that at $\Delta<\Delta_{\mathrm{cr}}$ the nodal points
are not accidental -- they are true symmetry-imposed $d-$wave nodal
points, protected by the fact that each is a Dirac point with a non-zero
winding number. Accordingly, because the damping near these new nodal
points is the same as near $d-$wave nodes on the FS in a conventional
case, all thermodynamic properties are the same as in a conventional
$d-$wave superconductor. Only in ARPES one can distinguish between
a conventional $d$-wave case with the nodes on the original FS and
the case when the nodes move away from the original FS due to the
presence of the inter-pocket pairing component. Still, this is a non-trivial
effect as the shift in $\epsilon_{\mathbf{k}}$ in Eqs. (\ref{a_9})
and (\ref{a_10}) vanishes along the directions $\sin{2\theta}=0$
and in this respect is qualitatively different from the overall shift
of the FS due to a change of the chemical potential.

The authors of Ref. \cite{si} also argued that the presence of inter-pocket
pairing component eliminates the nodes on the electron FS near $Z$
point in K$_{x}$Fe$_{2-y}$Se$_{2}$ ($Z=(0,0,\pi)$). We, on the
contrary, argue that this is not so, because the $Z$-pocket is made
out of single $d_{xy}$ orbital, with negligibly small admixture of
$d_{xz}$ and $d_{yz}$ orbitals, which at $Z$ point are located
far way from the chemical potential. In this situation, the nodes
should remain, if the pairing symmetry is $d-$wave. Moreover, the
displacement of the nodes from the FS is negligibly small, even if
the pocket itself is tiny, because the displacement is determined
by the ratio of the small $\Delta$ and the large distance between
the energies of $d_{xy}$ and other orbitals at $Z$.

\section{Conclusions}

In this work we analyzed the $d-$wave gap structure of multi-orbital
FeSC, as several experimental and theoretical studies suggested that
such a state may be realized in materials with only hole-like or only
electron-like Fermi pockets. We showed that the common belief that
a $d-$wave gap must have nodes right on the Fermi surfaces located
at the center of the BZ is correct only if this Fermi surface is made
out of a single orbital, but it is not true if there is an even number
of pockets made out of different orbitals. In FeSCs, there are two
pockets made out of $d_{xz}$ and $d_{yz}$ orbitals. We argue that
symmetry-imposed $d-$wave nodal points near $\Gamma$-point remain,
at least near $T_{c}$, but are shifted away from the normal state
FSs into the momentum region between the pockets. Depending on the
magnitude of the gap, as compared to the relative radii of the two
Fermi surfaces, the $d_{x^{2}-y^{2}}$-wave nodal points either persist
down to $T=0$, or come closer with decreasing $T$ and merge and
annihilate at a finite $T<T_{c}$ via a Lifshitz transition. This
transition, in which the Dirac gap nodes annihilate, is analogous
to a transition from a 2D massless Dirac semi-metal to an insulator.
Because the electron pockets are small and centered at $(\pi,0)$
and $(0,\pi)$, they do not cross the diagonals of the Brillouin zone,
i.e. there are no $d-$wave gap nodes on these pockets as well. Thus,
a $d-$wave FeSC with two $d_{xz}/d_{yz}$ hole pockets and two electron
pockets may display a completely nodeless $d-$wave superconductivity.

Our results have important consequences for the experimental identification
of $d$-wave states in FeSC. In particular, the fact that ARPES does
not see nodes in the $d_{xz}/d_{yz}$ Fermi surface of KFe$_{2}$As$_{2}$
is, in principle, not inconsistent with a $d-$wave state. However,
based on the values of the gap and of the radii of the $d_{xz}/d_{yz}$
hole pockets extracted from ARPES, it is likely that the nodes are
still present, buy away from the FS. In this regard, only precise
measurements of the gap along the hole pockets made predominantly
out of a single orbital can qualitatively distinguish between nodal
$s^{+-}$ and $d-$wave states. In this regard, the observation of
a nodeless gap in K$_{x}$Fe$_{2-y}$Se$_{2}$ on a Z-pocket consisting
of a single orbital, provides strong evidence against a $d-$wave
state.

\section{Acknowledgements}

We are thankful to E. Berg, S. Borisenko, P. Coleman, L. Fu, P. Hirschfeld,
J. Schmalian, and Q. Si for useful discussions. This work was supported
by the Office of Basic Energy Sciences, U.S. Department of Energy,
under awards DE-SC0014402 (AVC) and DE-SC0012336 (RMF). O.V. was supported
by NSF grant DMR 1506756. The authors thank the hospitality of the
Aspen Center for Physics, where part of this work was performed. ACP
is supported by NSF grant PHY-1066293.


\begin{thebibliography}{10}
\bibitem{comm} The superconducting gap on this pocket is visible
in the data, and its angular variation can be inferred without additional
analysis.
\end{thebibliography}
\end{document}